\documentclass[12pt]{article}
\usepackage[utf8]{inputenc}
\usepackage[
backend=biber,
style=numeric-comp, 
sorting=none 
]{biblatex}
\usepackage{epsfig,amsfonts,amssymb,framed,amsmath,xcolor}
\usepackage{slashed}
\usepackage{ytableau}
\usepackage{hyperref}
\usepackage{mathtools}
\usepackage{cancel}
\input epsf.sty
\usepackage{hhline}
\usepackage{bm}
\usepackage{array,upgreek}

\usepackage{ytableau}
\usepackage{braket}

\usepackage[margin=2cm]{geometry}
\newcommand{\be}{\begin{eqnarray}}
	\newcommand{\ee}{\end{eqnarray}}   

\newcommand{\bi}{\begin{itemize}}
	\newcommand{\ei}{\end{itemize}}

\newcommand{\bse}{\begin{subequations}}
	\newcommand{\ese}{\end{subequations}}

	\newcommand{\ul}{\underline{l}}

	\newcommand{\del}{\partial}
	\newcommand{\eps}{\epsilon}
	\newcommand{\eq}[1]{
		\begin{align}
			#1
		\end{align}
	}

	\newcolumntype{P}[1]{>{\centering\arraybackslash}p{#1}}
	
	\newcommand{\Z}{ \mathbb{Z} }

	\newcommand{\oh}{ {\frac{1}{2}} }
	\newcommand{\uV}{ {\underline V}}
\newcommand{\uchi}{ {\underline \chi}}
	\newcommand{\uA}{{\underline A}}
	\newcommand{\uL}{{\underline L}}
	\newcommand{\uk}{{\underline p}}
	\newcommand{\up}{{\underline p}}
	
	\newcommand{\ualpha}{{\underline \alpha}}

    \newcommand{\cL}{{\cal L}}

    \newcommand{\cN}{{\cal N}}

	\newcommand{\sdap}{ {\sqrt{2\alpha'}} } 
	\newcommand{\shap}{ {\sqrt{\frac{\alpha'}{2}}} }
    \newcommand{\hap}{\frac{\ap}{2}}

	\newcommand{\ap}{{\alpha'}}
    \newcommand{\azp}{\ualpha_0^+}
    \newcommand{\stap}{\sqrt{\frac{2}{\ap}}}
	\newcommand{\ishap}{  } 
	\newcommand{\zbar}{{\bar z}}

\begin{document}

		\baselineskip 24pt
		
		\begin{center}
			{\Large \bf }
			The Spectrum-generating Algebra for Bosonic Strings in a Linear Dilaton Background
		\end{center}
		
		\vskip .6cm
		\medskip
		
		\vspace*{4.0ex}
		
		\baselineskip=18pt

		\begin{center}
			{\large 
				\rm  Dripto Biswas$^a$}
		\end{center}

		\vspace*{4.0ex}
		\centerline{ \it \small $^a$ Dipartimento di Fisica, Universit\`{a} di Torino, 
			and I.N.F.N., Sezione di Torino}
		
		\centerline{ \it \small 
			Via P.\ Giuria 1, I-10125 Torino}

		\vspace*{1.0ex}
		\centerline{\small E-mail: dripto.biswas@unito.it}
		
		\vspace*{5.0ex}

		\centerline{\bf Abstract} \bigskip
		We extend the systematic construction of bosonic DDF operators to the light-like linear dilaton background to investigate how higher-spin string states behave beyond flat spacetime. Using previous results, we show that the spectrum-generating algebra is isomorphic to the flat spacetime case up to a few subtleties. This extension provides a controlled setting to explore higher-spin massive string interactions in a nontrivial yet exactly solvable string background. Most of the derivations lead to expressions very similar to those in the flat background, with the expected modifications appearing quite naturally in the spectrum and the deformed momentum-(non)conserving delta function.

		\vfill

		\vfill \eject

		\baselineskip18pt
		
		\tableofcontents

		\setcounter{section}{0}
		
\section{Introduction}
\label{sec:intro}

The study of higher-spin vertex operators in string theory represents one of the most technically challenging and theoretically rich areas in modern theoretical physics. These operators are fundamental to understanding the infinite tower of massive states that characterizes the spectrum of string theory. Although considerable progress has been made in the construction of such operators in flat spacetime backgrounds, their formulation in curved backgrounds remains an active frontier of research~\cite{GiveonLinearDilaton,AdS3Dilaton,ContinuumLinearDilaton,BarsHighSpin}.

Another reason why higher-spin string interactions are of significant interest is due to the Horowitz-Polchinski-Susskind black hole/string correspondence principle \cite{Susskind:1993ws,HorowitzPol1997,DAMOUR200093,Veneziano_2013,Chen2023,susskind2021note}, which suggests that perturbative string states may collapse into black holes when the closed string coupling $g_s$ reaches a critical threshold, $g_s  N^{1/4}\sim 1$, where $N$ represents the string’s excitation level \cite{HorowitzPol1997}. At this transition point, the black hole horizon area is of the typical size of the excited string, the Hawking temperature
matches the Hagedorn temperature,
and most crucially, the black hole entropy becomes comparable to the logarithm of the string state degeneracy ($\sim \sqrt{N}$). This correspondence provided the first statistical mechanical interpretation of black hole entropy in terms of microscopic string degrees of freedom. 

The DDF construction (Del Giudice, Di Vecchia, and Fubini) provides a systematic framework for building physical string states that are manifestly BRST invariant and satisfy the Virasoro constraints~\cite{DDFOriginal,Delgiudice1970}. This approach offers significant advantages over direct construction methods, especially when dealing with highly excited massive states where ensuring BRST invariance becomes exponentially more complex~\cite{Pesando:2024lqa,BarsHighSpin,Markou:2023ffh}. The DDF operators generate the complete physical spectrum by acting on a tachyonic ground state with carefully chosen null momenta, automatically incorporating the correct polarization structure and gauge invariances. The DDF construction was implemented and used to compute scattering amplitudes in various contexts \cite{Bianchi:2019ywd,Rosenhaus:2021xhm,Firrotta:2022cku,Hashimoto:2022bll,Das:2023xge,Savic:2024ock,Firrotta:2024qel,Firrotta:2024fvi,Bhattacharya:2024szw,Biswas:2024mdu,Das:2025prw,Alessio:2025nzd,Pesando:2024lqa,Pesando:2025ztr}.

Linear dilaton backgrounds represent a non-trivial yet exact string solution (the beta function vanishes). Among more formal applications, the use of the linear dilaton potential to regularize divergent Feynman integrals in lightcone superstring field theory was carried out in \cite{IshibashiLCgauge}.

Our approach leverages the \textit{framed} DDF (FDDF) formalism~\cite{FramedDDF}, which improves flexibility in the choice of reference frames and polarizations, maintaining BRST invariance and the construction of physical states. This proves valuable in curved backgrounds, where appropriate reference vectors may be constrained by geometry~\cite{FramedDDF}. In particular, we generalize the FDDF construction to a non-trivial background — the light-like linear dilaton — and show that the operator algebra, conformal properties, and hermiticity carry over with only controlled modifications. We also show that the \textit{improved} Brower states are null on-shell and decouple from the `physical' (lightcone) spectrum.

The theoretical foundation relies on the well-established relationship between spacetime physics and worldsheet CFT~\cite{Polchinski,Zwiebach}. Higher-spin vertex operators must be consistent with worldsheet conformal symmetry and spacetime gauge invariance, which powerfully constrains their form~\cite{VasilievReview,BRSTreview,BarsHighSpin}.

The rest of the paper is organized as follows: in Sec. \ref{sec:LDB} we briefly introduce the linear dilaton setup and our notation (along with some pedagogical details in App. \ref{app:LDB_setup}). In Sec. \ref{sec:FDDF} we collect the main results listing the definitions of the FDDF operators and their associated conformal properties. The systematic construction of spectrum-generating operators for the open bosonic string is performed by carefully analyzing modifications due to the linear dilaton profile in the Apps. \ref{app:uA},\ref{app:algebraconf} and \ref{app:Derivation of hermiticity properties}. OPE techniques produce the required correlators, BRST cohomology ensures gauge invariance~\cite{BRSTreview,BRSTDilaton}, and DDF construction provides complete coverage of the physical spectrum. The forms of the transverse FDDF operators in this background are isomorphic to those in the flat background up to certain subtleties, whereas the longitudinal (Brower) operators explicitly depend on the dilaton gradient. Nevertheless, they produce null states when on-shell and in critical dimensions $d=26$, which decouple from the physical spectrum; this is demonstrated using a simple explicit example in Sec. \ref{sec:example}.
\section{The linear dilaton background}
\label{sec:LDB}
The bosonic string worldsheet action in a linear dilaton background is given by
\eq{
S_{L.D.} = \frac{1}{4 \pi \ap}\int d^2\sigma \sqrt{h}\left[h^{ab}\left(\del_a X^\mu \del_b X_\mu\right) + \ap \mathcal{R}\Phi(X)\right],
}
where $h$ is the intrinsic metric and $\Phi(X) = V \cdot X$ is the linear dilaton field and ${\cal{R}} \equiv R^{(2)}$ is the Ricci scalar on the worldsheet. At the quantum level, this is an exactly solvable CFT (since the beta functions vanish to all orders in $\ap$). It is still possible to use the symmetries of this action to fix $h$ to be flat upto a \emph{boundary term} (see Appendix \ref{app:LDB_setup} for details on the classical setup). 

The holomorphic part of the stress tensor in this background is given by
\eq{
{\cal{T}}(z) := -\frac{1}{\ap} :\del L \del L:(z) + V \cdot \del^2 L(z),
\label{eq:stresstensor}
}
where $L(z)$ is the chiral left-moving, holomorphic component of the full string solution. The central charge of the CFT is \footnote{Since the linear dilaton does not affect the ghost action, the ghost CFT central charge is still $-26$.}
\eq{
c = d +6 \ap V^2,
}
where $d$ is the number of spacetime dimensions.
The modified Virasoro generators follow from \eqref{eq:stresstensor}:
\eq{
{\cL}_m &= \oint \frac{dz}{2\pi i}z^{m+1} {\cal T}(z) \nonumber \\
= \frac{1}{2}&:\sum_{n \in \mathbb{Z}}\alpha^\mu_{m-n}\alpha_{\mu n}: + i\shap(m+1) V^\mu \alpha_{\mu m}.
\label{eq:VirasoroGenLDB}
}
Since the dilaton is a topological effect, the string OPEs remain unchanged:
\eq{
L^\mu(z)L^\nu(w) =  -\frac{\ap}{2}g^{\mu\nu}\ln(z-w).
}
It can be shown that the correct (un-integrated) tachyon and photon vertex operators (in the `$-1$' ghost sector) in this background are given by \cite{Dodelson,CTchan}
\eq{
\mathcal{V}_T(x; k_T)
  &=
  c(x) : e^{i k_{T \mu } X^\mu(x, \bar x) }:
  ~~~~\mbox{with } \ap g^{\mu\nu} (k_{T \mu } k_{T \nu } + 2 i V_\mu k_{T\nu} )= 1
  \nonumber\\
  &=
  c(x) : e^{2 i k_{T \mu } L^\mu(x)} : ~~~~\mbox{when }x>0
  ,
  \nonumber\\
  \mathcal{V}_A(x; k, \epsilon)
  &=
  c(x)
  : \epsilon_\mu \partial_x X^\mu(x, \bar x) e^{i k_{\mu } X^\mu(x, \bar x) }:
  ~~~~\mbox{with }
  g^{\mu\nu} (k_{\mu } k_{\nu } + 2 i k_\mu V_\nu)= g^{\mu\nu} (k_{\mu } + 2 i V_\mu)\epsilon_\nu = 0  
  \nonumber\\
  &=
  c(x)
  : 2 \epsilon_\mu \partial_x L^\mu e^{2 i k_{ \mu } L^\mu(x)} :
  ~~~~\mbox{when }{x>0}
  ,
  \label{eq:tacpholindil-1}
}
where, the momenta in the linear dilation background satisfy the modified mass-shell conditions \footnote{In terms of the $k$ momentum variables, the zero modes gives rise to the momentum non-conservation $\delta^D(\sum k + i \chi_M V)$ in scattering amplitudes, where $\chi_M$ is the Euler number of the worldsheet.}.

To avoid possible complications arising from the Liouville potential in non-critical dimensions, as well as conformal anomalies, we restrict our computations to the case of a light-like linear dilaton $V^2 = 0$. 

\section{The \textit{framed} DDF approach}
\label{sec:FDDF}
In \cite{FramedDDF} we introduced the vielbein $E_\mu^{\underline{\mu}}$ and its inverse $E_\nu^{\underline{\nu}}$ with the property
\eq{
E_\mu^{\underline{\mu}} E_\nu^{\underline{\nu}} \eta_{\underline{\mu} \underline{\nu}}
= g_{\mu\nu}
,
\label{eq:framedef}
}
where $g_{\mu\nu}$ is the conformally flat metric (pushing the curvature due to the dilaton to $\infty$) \footnote{See Appendix A of \cite{Pei-MingHo_2008}  for a detailed path-integral computation.} appearing in the string action in the conformal gauge, and $\eta_{\underline{\mu} \underline{\nu}}$ is a dual flat metric. 
\subsection{Explicit form of FDDF operators in light-like linear dilaton background}
\label{sec:FDDF_LDB}
The underlying principle of the DDF construction is the explicit realization of the embedding structure $SO(d-2) \subset ISO(d-2) \subset SO(d-1,1)$ \cite{Goddard1972,NicolaiGebert_1997,DHOKER198790,HORNFECK1987189}.

The usual DDF construction begins with the choice of a Lorentz invariant vacuum (the tachyon for bosonic strings) corresponding to the Lorentz group $SO(d-1,1)$. 
This is followed by the choice of a null reference vector $q$ that corresponds to fixing a representation of the affine Euclidean subgroup $ISO(d-2) \subset SO(d-1,1)$, i.e., fixing the light-cone.
Finally, transverse polarization vectors, projectors, and DDF operators explicitly realize the subgroup $SO(d-2)$ that acts on the true physical degrees of freedom \footnote{For an introduction, see any standard text on string theory, for e.g. GSW Vol. I \cite{Green_Schwarz_Witten_2012}.}. The above steps are cleanly incorporated into the FDDF approach \cite{FramedDDF}. The associated tachyonic ground state can also be decoupled in the FDDF construction, allowing for off the mass-shell string computations using Mandelstam maps \cite{Biswas:2024epj}. 

Although the choice of $q$ (or the frame $E$) is completely arbitrary in flat space-time, it turns out that there is a `restriction' when $V \neq 0$. This is because there is a `preferred' affine direction in this case. The consistent choice is given by $E_\mu^{\underline{+}} = \lambda V_\mu;\, \lambda \neq 0$. From the above discussion, this is equivalent to choosing a certain `$V$-dependent' representation of $ISO(d-2)$.

We explicitly define the transverse FDDF operators in this background as
\eq{
\mathcal{\uA}^i_n := i\sqrt{\frac{2}{\ap}}\oint_{z=0} \frac{dz}{2\pi i}:\del \uL^i e^{\frac{in \uL^+(z)}{\ap \up_0^+}}:
\label{eq:DefDDFtrans}
}
where $\up_0^+ \neq 0$ is the corresponding null component of the zero-mode momentum operator and all underlined quantities are in the dual representation, for example $\uL^i = E_\mu^{\underline{i}}L^\mu,\, \uL^+ = E_\mu^{\underline{+}}L^\mu$, and so on. Although the form is exactly similar to the transverse FDDF operators in the $V=0$ case, we emphasize the additional constraints in this case: $\uV^i = \uV^+ = 0$. Note that $E^{\underline{i}}_\mu V^\mu = E^{\underline{+}}_{\mu}V^\mu = 0$, gives exactly $d-1$ constraints corresponding to the difference of symmetry generators of $SO(d-1,1)$ and $ISO(d-2)$. 

All quantities in the transverse and `$+$' null directions are the same as the corresponding flat space quantities in this restricted class of dual frames.
\subsection{The longitudinal (Brower) DDF operators - a difference}
Since the DDF construction is covariant (although not manifestly so), one also defines the longitudinal operators (see Appendices \ref{app:uA} and \ref{app:algebraconf} for related computations),
\eq{
{\cal{\uA}}^{-}_m = 
  i \sqrt{\frac{2}{\ap}}
  \oint_{z=0} \frac{d z}{ 2\pi i}
  :
  \left[
    \partial_z \uL^-(z)
    -
    \left(i\frac{m}{4 \up_0^+}
         +\ap \frac{\uV^-}{2}\right)\frac{\partial^2_z \uL^+}{\partial_z \uL^+}       
    \right]
    e^{i m \frac{ \uL^+(z) }{\ap \up^+_0} } :
    ,    
    \label{eq:diffA-}
}
The difference with respect to the flat space analog (with $\uV^-=0$) can be interpreted \textit{naïvely} \footnote{One could argue that the $V-$dependence could have been implicit just like the transverse ${\cal{\uA}}^i_n$. The form in \eqref{eq:diffA-} follows from the explicit conformal calculations.} as follows: the longitudinal DDF operators are representation-changing (as well as level-changing) operators (see \cite{NicolaiGebert_1997} for formal arguments and \cite{FramedDDF} for an explicit example at the string excitation level $N=1$). Since consistent representations (labeled by frames $E$) in a linear dilaton background are themselves $V-$dependent, it follows that representation-changing operators are also $V-$dependent.

We further define the \textit{improved} Brower operator in a linear dilaton background,
\begin{equation}
   {\widetilde    {\cal{\uA}}}^-_m(E)
   =
   {{\cal{\uA}}}^-_m(E)
   -
   \frac{1}{\ualpha_0^+} \tilde{\cL}_m(E)
   -
   \frac{d-2}{24}
   \frac{1}{\ualpha_0^+}\,
   \delta_{m,0} 
   ,
   \label{eq:diffAtilde-}
\end{equation}
where we have defined the Sugawara operators (replacing $\ualpha \rightarrow {\cal\uA}$) as
\begin{align}
\tilde{\cL}_m(E)
=
&
\frac{1}{2}
\sum_{j=2}^{D-1} \sum_{l\in \Z}: {\cal{\uA}}^j_l(E)\, {\cal{\uA}}^j_{m-l}(E) : - i\sqrt{\frac{\ap}{2}}(m+1)\uV^- \ualpha_0^+\delta_{m,0}
.
\end{align}
The second term in the expression above (from ${\cal \uA}^+_m$) arises from the $V$-dependent term in ${\cal T}(z)$.
\subsection{Algebra and conformal properties}
\label{sec:algebraconf}
The algebra satisfied by the FDDF operators are, again, isomorphic to the flat space analogs:
\begin{align}
  [{\cal{\uA}}^i_m(E_V),\, {\cal{\uA}}^j_n(E_V)]
  &= m\, \delta_{m+n,0} \delta^{i j}
  ,
  \label{eq:DDFalgebra}
  \\
  [{\cal{\uA}}^i_m(E_V),\, \ualpha_0^+{\cal{\uA}}^-_n(E_V)]
  &=
   m
  \,
  {\cal{\uA}}^i_{m+n}(E_V)
  \label{eq:Ai A- algebra}
  \\
   [\ualpha_0^+ {\cal{\uA}}^-_m(E_V),\, \ualpha_0^+ {\cal{\uA}}^-_n(E_V)]
   &=
   (m-n)\, \ualpha_0^+ {\cal{\uA}}^-_{m+n}(E_V)
   +
   2 m^3\,
   \delta_{m+n, 0}
   ,
  \label{eq:A- A- algebra}
\end{align}
and
\begin{align}
  [{\cal{\uA}}^i_m(E_V),\, \ualpha_0^+ \tilde {\cal{\uA}}^-_n(E_V)]
  &=
0 
  \nonumber\\
   [\ualpha_0^+ \tilde {\cal{\uA}}^-_m(E_V),\, \ualpha_0^+ \tilde {\cal{\uA}}^-_n(E_V)]
   &=
   (m-n)\, \ualpha_0^+ \tilde {\cal{\uA}}^-_{m+n}(E_V)
   +
   \frac{26 - d}{12}
   m^3\,
   \delta_{m+n, 0}
   .
  \label{eq:DDFalgebra_tilde}
\end{align}
We have emphasized that the choice of frames $E_V$ is dependent on the dilaton potential $V \neq 0$ and furnish representations of $ISO(d-2)$ as opposed to arbitrary frames $E$ which label representations of the Lorentz group $SO(d-1,1)$. However, as shown in \cite{FramedDDF}, the algebra is independent of the choice of $E$ (and therefore by extension, of the choice of $E_V$) \footnote{We still perform the explicit computations in Appendix \ref{app:algebraconf} as it is a non-trivial exercise for the modified longitudinal operators and Virasoro generators ${\cal{\uL}}_n$ in this background.}. We shall henceforth drop the explicit frame dependence for most of the following computations, keeping in mind the context above for a linear dilaton background.

\subsubsection*{Conformal properties}
It follows from the previous discussion straightforwardly that
\eq{
[{\cal{L}}_n,{\cal \uA}^i_m] = [L_n, {\cal \uA}^i_m] = 0,
}
where $L_n$ is the Virasoro generator for $\uV = 0$. This is because the OPE of $\uV^- \del^2 \uL^+$ in ${\cal{T}}(z)$ with the $\uL^i, \uL^+$ components of the string field in ${\cal{\uA}}^i_m$ vanish trivially. The second equality may be proved as
\eq{
  [L_n,{\cal \uA}^i_m]=&
    i \sqrt{\frac{2}{\ap}}
  \frac{-2}{\ap}
  \oint_{w=0}  
  \oint_{z=w}
  z^{n+1}
  :
  e^{\delta\cdot \partial \uL(z)}
  e^{\underline\epsilon \cdot \partial \uL(w) + i \uk \cdot \uL(w)}
  :
  e^{
    -
    \hap \frac{\delta\cdot \epsilon}{(z-w)^2}
    -
    \hap
    \frac{i \delta\cdot \uk}{z-w}
  } \Bigg \lvert_{\delta^2, \eps_i, \up_+}
  \nonumber \\
  =&
  i \sqrt{\frac{2}{\ap}}
  \oint_{w=0}  
  :
  \left[
    \underline{\epsilon} \cdot \partial( w^{n+1} \partial \uL(w) )
    +
    w^{n+1}
    \underline\epsilon \cdot \partial \uL(w)
    i \uk \cdot \partial \uL(w)
    \right]
  e^{ i \uk \cdot \uL(w)}
  :  
  \nonumber\\
  =&
  i \sqrt{\frac{2}{\ap}}
  \oint_{w=0}  
  :
\partial
  \left[
    w^{ \ap \uk \cdot \up_0 + n + 1}
    \dots
    \right] = 0
  .
  \label{eq:LA_comm}
}
The importance of $1/\up_0^+$ in the exponential of the (F)DDF operator is clear in the last line as this removes the cut inside the total derivative.
\subsubsection*{Hermiticity properties}
The hermiticity properties of the framed operators are the expected ones and the same as in the flat spacetime case.
\begin{equation}
\left[ {\cal\uA}^i_m(E) \right]^\dagger
=
{\cal\uA}^i_{-m}(E)
~~,~~
\left[ {\cal\uA}^-_m(E) \right]^\dagger
=
{\cal\uA}^-_{-m}(E)
~~,~~
\left[ \tilde {\cal\uA}^-_m(E) \right]^\dagger
=
\tilde {\cal\uA}^-_{-m}(E)
.
\label{eq:hermieqs}
\end{equation}  
However, in the case of ${\cal\uA}^-_{-m}(E)$, the exact computation is much more involved. There is an `internal' breaking of hermiticity in the linear dilaton background that is neatly re-packaged using the shifted string coordinate $\uchi(z) = \uL(z) + \ap \uV \ln(z)$ as shown in detail in Appendix \ref{app:Derivation of hermiticity properties}. However, the final result is exactly as in \eqref{eq:hermieqs}. This allows us to almost directly \footnote{The zero-mode integral in the `-' direction giving the modified momentum(non)-conservation will be important to obtain `on-shell' results in this background.} use the scattering computations in \cite{Biswas:2024mdu}.
\section{An explicit example: level $N=1$ DDF state}
\label{sec:example}
In this section, we look at the lowest excited DDF state in the modified spectrum and confirm that it indeed satisfies the modified Virasoro condition. One can directly compute using the definition \eqref{eq:DefDDFtrans} that
\eq{ 
    {\cal\uA}^i_{-1} \ket{\up_T}
    =
    \left[
      \ualpha^i_{-1}
    -
    \frac{\uk^i }{ \uk^+ } \ualpha^+_{-1}
    \right]
    \ket{\uk_{T + 1}^-, \uk_{T}^+, \uk_{T }^i} 
, 
\label{eq:N1DDDFtransex}
}
where, the (only) shifted null momentum is defined as $\up_{T+N}^- = \up_{T}^- +\frac{N}{2\ap \uk^+}$. Comparing \eqref{eq:N1DDDFtransex} with the covariant level $N=1$ state: $|\epsilon, p\rangle
  = \underline{\eps}_\mu \ualpha^\mu_{-1} | \uk_\nu\rangle
  = \epsilon_\mu \alpha^\mu_{-1} | p_\nu\rangle$, we get
\eq{
\eps^{(i)}_\mu = E^{\underline{i}}_{\mu} - \frac{E^{\underline{i}}_{\rho}\,p^\rho}{V \cdot p}V_\mu = \Pi^{\underline{i}}_\mu(E_V) ,
\label{eq:transproj}
}
where, we have made explicit the restricted choice of vielbein in the linear dilaton background ($E^{\underline{+}}_\mu \rightarrow \lambda V^\mu$). $\Pi^{\underline{i}}_\mu(E_V)$ is the modification of the usual transverse projector that appears in the flat spacetime case.

Finally, using \eqref{eq:VirasoroGenLDB} the only Virasoro condition at level $N=1$ is given by,
\eq{
{\cal L}_{1}\ket{\eps, p} = \eps \cdot (p + 2 i V) = \Pi^{\underline{i}}_\mu p^\mu = 0,
}
where, in evaluating the last expression, we used \eqref{eq:transproj} along with $V^2= \uV^i = 0$.
\subsection{A comment on the role of Brower operators}
The role of (longitudinal) Brower operators as representation-changing operators was pointed out in \cite{NicolaiGebert_1997}. In the $\uV^-=0$ case, an explicit example was also computed in \cite{FramedDDF} to show that these states become null on-shell and in critical dimension, thereby decoupling from the `physical' spectrum. In the linear dilaton background, this is still the case as demonstrated in the following simple example:
\subsubsection*{Level $N=1$ Brower state}
Instead of writing the full \textit{improved} Brower state at level $N=1$ (which follows from the definitions of the components involved), we simply compute the norm using the algebra in \eqref{eq:DDFalgebra_tilde}:
\eq{
  \langle \uk_T|  \tilde{\cal \uA}^-_{1}   \tilde{\cal \uA}^-_{-1} | \ul_T \rangle
  &=
  \langle \uk_T|
  \left[
  2 \frac{1}{\ualpha^+_0 } \tilde{\cal \uA}^-_{0}
  +
  \frac{26-d}{12} \frac{1}{ (\ualpha^+_0)^2 }
  \right]
  | \ul_T \rangle  \nonumber \\
  =
  - &\frac{-2 \ul^-_{T+1} \ul^+ +  \vec \ul\,^2 + 2 i \uV^- \ul^+}{ 2 (\ul^+)^2}
  \delta(\uk-\ul - i \uV)
  = - 
  \frac{\ul \cdot (\ul + 2 i \uV)}{2(\ul^+)^2}  \delta(\uk-\ul - i \uV)
  \label{eq: vev tAm0 is zero}
}
where $\ul_{T+1}^- = \ul_T^- + \frac{1}{2\ap \ul_T^+};\,\ul_T^+ = \ul^+;\,\ul_T^i = \ul^i$ and we have set $d=26$. The deformed delta function arises from the modified inner product in the presence of a dilaton
\eq{
\langle k'|k\rangle = \delta(k' - k - iV),
}
where $p_0\ket{k} = k \ket{k};\, \bra{k'}p_0 = \bra{k'-iV}(k'-iV)$ \footnote{In the `framed' notation, $\up_0^-$ is non-Hermitian - this is important in the derivation of the correct Hermiticity properties in Appendix \ref{app:Derivation of hermiticity properties}, as well as to get the modified momentum conservation in the Reggeon.}. 

This vanishes from the modified mass-shell condition of a photon in a linear dilaton background (see \eqref{eq:tacpholindil-1}). The above computation can be extended to the most general FDDF states given by
\eq{
\prod_{m=1}^\infty ({\cal\uA}^i_{-m}(E_V))^{N^i_m}
(\tilde {\cal \uA}^-_{-m}(E_V))^{N_m}
c_{-1} \ket{\uk_T}
,
}
which are physical since (for $n \ge 1$)
\eq{
\cL_{n} \prod_{m=1}^\infty ({\cal\uA}^i_{-m}(E_V))^{N^i_m}
(\tilde {\cal \uA}^-_{-m}(E_V))^{N_m}
c_{-1} \ket{\uk_T} = \prod_{m=1}^\infty ({\cal\uA}^i_{-m}(E_V))^{N^i_m}
(\tilde {\cal \uA}^-_{-m}(E_V))^{N_m}
c_{-1} \cL_n \ket{\uk_T} = 0,
}
and null on-shell (i.e. BRST exact) in $d=26$ for $N_m, \neq 0$.

Therefore, the \textit{improved} Brower states are null on-shell and decouple just as in the flat space case. 

However, as pointed out in \cite{Pei-MingHo_2008}, the massless sector in this background contains discrete states ($\ket{D^i},\ket{D^+}, \ket{D^-}$ in their notation), which are not spanned by the DDF operators. These states are not in the accessible Fock space of the (F)DDF operators, since they have $\up_0^+\ket{D^\mu} = 0$ irrespective of the choice of the vielbein $E_V$. Although the coupling of $\ket{D^-}$ to physical (massless) states is non-zero \cite{Pei-MingHo_2008}, they do not break unitarity since the phase space available to discrete states has measure zero.
\section{Summary and concluding remarks}
In this paper, we provided a careful extension of the (F)DDF approach \cite{FramedDDF} to higher-spin vertex operators to a light-like linear dilaton background. In Sec. \ref{sec:algebraconf}, we showed that the spectrum-generating algebra is isomorphic to the flat ($V=0$) case even though the longitudinal (\textit{improved} Brower) operators ${\cal\uA^-} (\tilde {\cal \uA}^-)$ contain $V$-dependent deformations in their explicit representations \eqref{eq:diffA-} and \eqref{eq:diffAtilde-}. We also showed using the above algebra, in Sec. \ref{sec:example}, that the null states generated by the \textit{improved} Brower operators indeed decouple from the physical spectrum in the linear dilaton background as well. 

Furthermore, because of the conformal and Hermiticity properties in Sec. \ref{sec:algebraconf} (details in Appendices \ref{app:algebraconf} and \ref{app:Derivation of hermiticity properties}), we can straightforwardly extend the Sciuto-Della Selva-Saito (SDS)~\cite{Sciuto1969,Selva1970} inspired DDF Reggeon construction in \cite{Biswas:2024mdu} to the linear dilaton background, with the only modification being the lightcone momentum-(non)conservation given by
\eq{
\delta\left(\sum_{a=1}^M \up_a^- - i \uV^-\right)\delta\left(\sum_a \up_a^+\right)\delta^{(d-2)}\left(\sum_a \up_a^i\right).
}

The effect of modified kinematics on massive string scattering amplitudes (using DDF), in particular the changes, if any, to the \emph{chaotic} behavior pointed out in~\cite{Gross:2021gsj} and related changes in the measures~\cite{Bianchi:2022mhs,Bianchi:2023uby}, as well as classical string profiles~\cite{Das:2025prw} is left for future work.

It would also be interesting to find an explicit DDF construction for arbitrary linear dilaton backgrounds in the presence of the Liouville potential. 
\section*{Acknowledgments}
The author thanks Igor Pesando for helpful discussions and comments during the preparation of this draft.
\appendix
\section{The setup: linear dilaton background}
\label{app:LDB_setup}
The Polyakov action for a bosonic string propagating in a background metric 
\( G_{\mu\nu}(X) \), antisymmetric tensor \( B_{\mu\nu}(X) \), and dilaton field \( \Phi(X) \) is
\begin{equation}
S = \frac{1}{4\pi \alpha'} \int d^2\sigma \sqrt{h} \left[
h^{ab} G_{\mu\nu}(X) \, \partial_a X^\mu \partial_b X^\nu
+ \epsilon^{ab} B_{\mu\nu}(X) \, \partial_a X^\mu \partial_b X^\nu
+ \alpha' R^{(2)} \Phi(X)
\right].
\end{equation}

In the \emph{linear dilaton background}, we take
\begin{equation}
G_{\mu\nu} = \eta_{\mu\nu}, \qquad B_{\mu\nu} = 0, \qquad \Phi(X) = V_\mu X^\mu,
\end{equation}
where \( V_\mu \) is a constant vector, often referred to as the \emph{dilaton gradient}.

\subsection*{Explicit symmetries of the linear dilaton Action}

\subsubsection*{Worldsheet symmetries}

The action is classically invariant under the usual Polyakov symmetries:

\paragraph{(a) Worldsheet diffeomorphism invariance:}
\[
\sigma^a \to \sigma'^a(\sigma),
\]
with \( h_{ab} \) transforming as a tensor and \( X^\mu \) as scalars.

\paragraph{(b) Weyl invariance:}
\[
h_{ab} \to e^{2\omega(\sigma)} h_{ab}, \qquad X^\mu \to X^\mu.
\]
Under a Weyl transformation, the scalar curvature changes as
\[
\sqrt{h} R^{(2)} \to \sqrt{h} (R^{(2)} - 2\nabla^2 \omega),
\]
so that the action shifts by
\begin{equation}
\delta_\omega S = -\frac{1}{2\pi} \int d^2\sigma \sqrt{h} \, \nabla^2 \omega \, \Phi(X)
= -\frac{1}{2\pi} \int_{\partial\Sigma} ds \, n^a \partial_a \Phi(X)\, \omega.
\end{equation}
Hence, the Weyl invariance holds \emph{up to a boundary term}, corresponding to a \emph{boundary charge}.

Thus, classically, the action is invariant under both worldsheet diffeomorphisms and Weyl rescalings, up to boundary contributions.

\subsection*{Target space symmetries}

Since \( G_{\mu\nu} = \eta_{\mu\nu} \) and \( B_{\mu\nu} = 0 \), the kinetic term
\begin{equation}
S_X = \frac{1}{4\pi \alpha'} \int d^2\sigma \sqrt{h}\, h^{ab} \partial_a X^\mu \partial_b X_\mu
\end{equation}
is invariant under the full Poincaré group:
\[
X^\mu \to \Lambda^\mu{}_\nu X^\nu + a^\mu.
\]

The dilaton coupling term,
\begin{equation}
S_\Phi = \frac{\alpha'}{4\pi} \int d^2\sigma \sqrt{h}\, R^{(2)} V_\mu X^\mu,
\end{equation}
modifies these symmetries:

\paragraph{(a) Translations:}
Under \( X^\mu \to X^\mu + a^\mu \),
\begin{equation}
S_\Phi \to S_\Phi + \frac{\alpha'}{4\pi} a^\mu V_\mu \int d^2\sigma \sqrt{h}\, R^{(2)}.
\end{equation}
Since
\[
\int d^2\sigma \sqrt{h}\, R^{(2)} = 4\pi \chi(\Sigma)
\]
is topological (Euler characteristic), the shift
\(
\delta S = \alpha' a^\mu V_\mu \chi(\Sigma)
\)
is a constant and does not affect the equations of motion. Thus, \emph{translations remain a symmetry up to a topological term}.

\paragraph{(b) Lorentz transformations:}
\( X^\mu \to \Lambda^\mu{}_\nu X^\nu \) preserves \( S_\Phi \) only if \( V_\mu \) is invariant:
\[
V_\mu \to V_\nu (\Lambda^{-1})^\nu{}_\mu = V_\mu.
\]
Hence, only the subgroup of the Lorentz group that leaves \( V_\mu \) fixed remains a symmetry—the \emph{stabilizer subgroup} of \( V_\mu \) (for null $V$, this is exactly the associated Euclidean subgroup $ISO(d-2)$ appearing in the embedding structure in Sec. \ref{sec:FDDF_LDB}).

\begin{table}
\centering
\begin{tabular}{|c|c|c|}
\hline
\textbf{Type} & \textbf{Symmetry} & \textbf{Status in Linear Dilaton} \\
\hline
Worldsheet & Diffeomorphism & Exact \\
 & Weyl invariance & Up to boundary term \\
\hline
Target-space & Translations & Up to topological term \\
 & Lorentz transformations & Broken to stabilizer of \(V_\mu\) \\
\hline
\end{tabular}
\caption{Summary of explicit symmetries}
\end{table}

\subsection*{Why a flat target-space metric (still) works}

The combined worldsheet diffeomorphism and Weyl invariance allow us to fix the worldsheet metric to the conformal gauge:
\begin{equation}
h_{ab} = e^{2\rho(\sigma)} \eta_{ab}.
\end{equation}
The Weyl symmetry removes the factor \( e^{2\rho} \), except for a possible boundary charge proportional to
\(
\int R^{(2)} \Phi(X),
\)
which is purely topological.

Moreover, the conformal invariance conditions (vanishing of the beta functions) are satisfied in the linear dilaton background if
\begin{equation}
D + 6\alpha' V^2 = 26,
\end{equation}
which ensures \emph{quantum Weyl invariance} even though the dilaton has a nontrivial gradient.

Therefore, we can consistently choose a flat target-space metric \( G_{\mu\nu} = \eta_{\mu\nu} \); the dilaton gradient \( V_\mu \) compensates for the central charge deficit that would otherwise require a curved background.
\section{${\cal{\uA}}^-$ and $\tilde {\cal{\uA}}^-$ operators}
\label{app:uA}
We collect here the derivation of the improved longitudinal operator ensuring absence of cubic poles in OPEs with the modified stress tensor ${\cal{T}}(z)$.
Just as in the flat space case, the naive version of the longitudinal DDF operator can be written as,
\eq{
 &{\widehat    {\cal{\uA}}}^-_m
  =
  i \sqrt{\frac{2}{\ap}}
  \oint_{z=0} \frac{d z}{ 2\pi i}
  : \partial_z \uL^-(z) e^{i m \frac{ \uL^+(z) }{\ap \up^+_0} } :
\label{eq:Ahat}
}

To see the non-vanishing cubic pole, we use the modified Virasoro generators 
\eq{
{\cal L}_n = \oint \frac{dz}{2\pi i} z^{n+1}\left(-\frac{1}{\alpha'}:\del {\uL}(z)\del {\uL}(z): + \uV \cdot \del^2 \uL\right),
}
to obtain
\eq{
[{\cal L}_n,\widehat    {\cal{\uA}}^-_m] &= \left[\oint_z z^{n+1}\left(-\frac{2}{\alpha'}\right)e^{\delta \cdot \del \uL(z) - \hap \uV \cdot \del^2 \uL(z)}, i\stap \oint_w e^{\eps \cdot \del \uL(w) + i k \cdot \uL(w)}\right]\Biggm|_{\delta^2,\eps_-,k_+, \uV^-} \nonumber \\
= -i \frac{2}{\alpha'}\stap \oint_w \oint_{z=w}& z^{n+1} :e^{\delta\cdot \del \uL(z)- \hap \uV \cdot \del^2 \uL(z)} e^{\eps \cdot \del \uL(w) + i k \cdot \uL(w)}: e^{-\frac{\alpha'}{2}\frac{\delta \cdot \eps}{(z-w)^2}}e^{-i\frac{\alpha'}{2}\frac{\delta\cdot k}{(z-w)}}e^{-\frac{\ap^2}{2}\frac{\eps \cdot \uV}{(z-w)^3}}\Biggm|_{\delta^2,\eps_-,k_+, \uV^-} \nonumber \\
= i \stap \oint_w \oint_{z=w} z^{n+1}&\left[-i\frac{\alpha'}{2} \frac{(\delta\cdot \eps)(\delta \cdot k)}{(z-w)^3} + \frac{\delta\cdot\eps}{(z-w)^2}\delta\cdot\del\uL(z) \right. \nonumber \\
+& \left.i\delta \cdot \del\uL(z) \eps\cdot \del \uL(w) \frac{\delta \cdot k}{(z-w)} - \ap \frac{\eps \cdot \uV}{(z-w)^3}\right]e^{i k \cdot \uL(w)}\Biggm|_{\delta^2,\eps_-,k_+, \uV^-} \nonumber\\
}
Using $\delta_\mu \delta_\nu = \eta_{\mu\nu}$, $\eps\cdot k = -\frac{m}{\alpha' \up_0^+}$ and $k \cdot \uV = 0$ we get,
\eq{
[{\cal L}_n,\widehat    {\cal{\uA}}^-_m] = i\stap \oint_{w=0}&:\left[\eps \cdot \del(w^{n+1}\del\uL(w)) + i w^{n+1}\eps\cdot\del\uL(w) k \cdot\del\uL(w)\right]e^{i k \cdot \uL(w)}: \nonumber \\
+ &i\stap \oint_{w}\oint_{z=w}z^{n+1}\left(-i \frac{\alpha'}{2}\frac{\eps \cdot k}{(z-w)^3} - \ap \frac{\eps \cdot \uV}{(z-w)^3}\right)e^{i k\cdot \uL(w)} \nonumber \\
= &i\stap \oint_{z=0} \del^2(z^{n+1})\left[\frac{im}{4p_0^+} + \ap \frac{\uV^-}{2}\right]e^{\frac{im\uL^+(z)}{\alpha'\up_0^+}} + ...,
\label{eq:cubic_term}
}
where, ``..." denotes the usual total derivative terms which vanish. To remove this cubic pole contribution, we first compute (recall that only $\uV^- \neq 0$),
\eq{
\left[\del_w \uL^-, \del^2_z\uL^+/\del_z \uL^+\right]&= -\del_z \int_0^\infty \frac{d\xi}{\xi} 
  [\partial_w \uL^-,
  e^{- \xi \partial_z \uL^+}]
  \nonumber\\
=&\hap \del_z \int_0^\infty d\xi \frac{1}{(w-z)^2}e^{-\xi \del_z \uL^+} + :..:\nonumber \\
=&
  \hap\partial_z
  \left[
    \frac{1}{(w-z)^2}     \frac{1}{\partial_z \uL^+}
    \right]
  +
  :
  \partial_w \uL^-\,
  \frac{\partial^2_z \uL^+}{\partial_z \uL^+}
  :,
\label{eq:Ln_corr1}
}
where we used the OPE for $\del_w \uL^- \del_z \uL^+$ directly in the second step. Using \eqref{eq:Ln_corr1}, we get
\eq{
[{\cal L}_n,&\del^2_z\uL^+/\del_z \uL^+] = [L_n,\del^2_z\uL^+/\del_z \uL^+] = -\frac{2}{\alpha'}\oint_w\left[\del_w \uL^- , \frac{\del^2_z\uL^+}{\del_z \uL^+} \right]w^{n+1}\del_w \uL^+ \nonumber \\
&= \del_z \left[\del_w\left(w^{n+1}\del_w \uL^+\right)\Biggm|_{w=z} \frac{1}{\del_z \uL^+}\right] = \del^2_z(w^{n+1}) + \del_z\left(z^{n+1} \frac{\del^2_z\uL^+}{\del_z \uL^+}\right)
\label{eq:Ln_corr2}
}
Finally, we can compute
\eq{
\Bigg[{\cal L}_n,&
    i \ishap
    \oint_{z=0} \frac{d z}{ 2\pi i}
    \frac{\partial^2_z \uL^+}{\partial_z \uL^+}
    e^{i m \frac{ \uL^+(z) }{\ap \up^+_0} }
  \Bigg] = i\ishap \oint_z \left(\frac{\del^2_z \uL^+}{\del_z \uL^+}\Big[L_n,e^{i\frac{m \uL^+(z)}{\ap \up_0^+}}\Big] + \left[L_n,\frac{\del^2_z \uL^+}{\del_z \uL^+}\right]e^{i\frac{m \uL^+(z)}{\ap \up_0^+}}\right) \nonumber \\
&= i \ishap \oint_z \left(\del^2_z(z^{n+1})e^{i\frac{m\uL^+(z)}{\ap \up_0^+}} + \del_z\left(z^{n+1}\frac{\del^2_z \uL^+}{\del_z \uL^+}\right) + \frac{\del^2_z \uL^+}{\del_z \uL^+} z^{n+1}\del_z\left(e^{i \frac{m\uL^+}{\ap \up_0^+}}\right)\right) \nonumber \\
&= i\ishap\oint_z \del^2_z(z^{n+1})e^{i\frac{m\uL^+(z)}{\ap \up_0^+}} + i\ishap \oint_z \del_z\left(z^{n+1}\frac{\del^2_z \uL^+}{\del_z \uL^+}e^{i\frac{m\uL^+(z)}{\ap \up_0^+}}\right),
}
where the second term is zero since it does not contain any branch cuts and is a total derivative. Therefore, the longitudinal DDF operator in a light-like linear dilaton background, with the correct conformal properties (free of cubic pole contributions) is defined as,
\begin{equation}
{\cal\uA}^-_m
  =
  i \sqrt{\frac{2}{\ap}}
  \oint_{z=0} \frac{d z}{ 2\pi i}
  :
  \left[
    \partial_z \uL^-(z)
    -
    \left(i\frac{m}{4 \up_0^+}
         +\ap \frac{\uV^-}{2}\right)\frac{\partial^2_z \uL^+}{\partial_z \uL^+}       
    \right]
    e^{i m \frac{ \uL^+(z) }{\ap \up^+_0} } :
    ,    
\label{eq:A-_deriv}
\end{equation}
which satisfies $[{\cal L}_n, {\cal\uA}^-_m] = 0$ by construction.
\section{Derivation of algebra and conformal properties}
\label{app:algebraconf}
The ${\cal{\uA}}^i_m$ algebra can be obtained via the usual means. Let us  write
\begin{align}
{\cal{\uA}}^i_m =
  \cN
  \oint_{z=0} \frac{d z}{ 2\pi i}
  : \partial_z \uL^i(z) e^{i m \delta_+ \uL^+(z)} :
  ,
  \label{eq:DDFflat}
\end{align}
with
\begin{equation}
\cN = i \sqrt{\frac{2}{\ap}},~~~~
\delta_+ =  \frac{1}{\ap \up_0^+},
~~~~
(\mbox{when } \up^+_0\ne0)
\label{eq:N_delta_vals}
,
\end{equation}
then  we get,
\begin{align}
[{\cal{\uA}}^i_m, {\cal{\uA}}^j_n]
&=
\cN^2
\left[
\oint_{z=0, |z|>|w|} \frac{d z}{ 2\pi i}
\oint_{w=0} \frac{d w}{ 2\pi i}
-
\oint_{w=0} \frac{d w}{ 2\pi i}
\oint_{z=0, |z|<|w|} \frac{d z}{ 2\pi i}
\right]
\nonumber\\
&
R\left[
: \left( \partial_z \uL^i e^{i m \delta_+ \uL^+} \right)(z) :
: \left( \partial_w \uL^j e^{i n \delta_+ \uL^+} \right)(w) :
\right]
%
%
\nonumber\\    
&=
\cN^2
\oint_{w=0} \frac{d w}{ 2\pi i}
\oint_{z=w} \frac{d z}{ 2\pi i}
\Biggl[
-\hap \frac{ \delta^{i j} }{(z-w)^2} e^{i (m+n) \delta_+ \uL^+(w) }
\nonumber\\
&~~~~
-
\hap \frac{ \delta^{i j} }{(z-w)}
i \delta_+ m \partial_w \uL^+ e^{i (m+n) \delta_+ \uL^+(w) }
+\dots
\Biggr]
\nonumber\\
&=
\left( -\oh (\ap\cN)^2 \delta_+ \up^+_0 \right)
m \delta_{m+n,0} \delta^{i j}
.
\label{eq:AA_comm_norm}
\end{align}
Using the definitions above we can compute,
\eq{
[{\cal{\uA}}^i_m,\ualpha_0^+ {\cal{\uA}}^-_n] &= [{\cal{\uA}}^i_m,\ualpha_0^+ {\widehat    {\cal{\uA}}}^-_n] = -\frac{2}{\alpha'} \ualpha_0^+\oint_{w=0}\oint_{z=w} :e^{\delta\cdot\del_z\uL + iq\cdot\uL(z)}::e^{\gamma\cdot\del_w\uL + ik\cdot \uL(w)}:\Bigm|_{\delta_i,\gamma_-} \nonumber \\
= -\frac{2}{\alpha'} &\ualpha_0^+\oint_{w=0}\oint_{z=w} :e^{\delta\cdot\del_z\uL + iq\cdot\uL(z)} e^{\gamma\cdot\del_w\uL + ik\cdot \uL(w)}: e^{i\hap\frac{q\cdot\gamma}{(z-w)}}\Bigm|_{\delta_i,\gamma_-}\nonumber\\
=&-\frac{2}{\ap}\ualpha_0^+\oint_{w=0}\oint_{z=w}:\del_z \uL^i e^{i\frac{(m \uL(z) + n \uL(w))}{\ap \up_0^+}}: \left(-\frac{im}{\ap \up_0^+}\right)\hap\frac{1}{(z-w)} \nonumber \\
&\implies [{\cal{\uA}}^i_m,\ualpha_0^+ {\cal{\uA}}^-_n] = m {\cal{\uA}}^i_{m+n},
\label{eq:AiA-}
}
where, $\delta_i = 1, q_+ = m/(\ap \up_0^+), \gamma_- = 1, k_+ = n/(\ap \up_0^+)$ are the only non-zero components.

We now calculate the commutator,
\eq{
[\azp {\cal{\uA}}^-_m, \azp {\cal{\uA}}^-_n] = (\azp)^2([\widehat {\cal{\uA}}^-_m,\widehat    {\cal{\uA}}^-_n] - [\widehat {\cal{\uA}}^-_m,C_n] - [C_m,\widehat    {\cal{\uA}}^-_n] + [C_m,C_n]),
\label{eq:AmAm_full}
}
where, $C_m =i\stap \left(\frac{im}{4\up_0^+} +\ap \frac{\uV^-}{2}\right)\oint_{z=0} :\frac{\del^2_z \uL^+}{\del_z\uL^+} e^{\frac{im\uL^+}{\ap \up_0^+}}:$. We see that,
\eq{
[\widehat {\cal{\uA}}^-_m,\widehat    {\cal{\uA}}^-_n] =& -\frac{2}{\ap}\oint_{w=0}\oint_{z=w}:e^{\delta\cdot \del_z\uL + iq\cdot\uL(z)}e^{\eps\cdot\del_w\uL+ik\cdot\uL(w)}:e^{-i\frac{\delta\cdot k}{(z-w)}\hap}e^{i \frac{\eps\cdot q}{(z-w)}\hap}\Bigm|_{\delta_-,\eps_-}\nonumber\\
=-\frac{2}{\ap}\oint_{w=0}&\oint_{z=w}\left[-\frac{im}{\ap \up_0^+}\frac{\del_z\uL^-}{(z-w)}\hap + \frac{in}{\ap \up_0^+}\frac{\del_w \uL^-}{(z-w)}+\left(\hap\right)^2\frac{mn}{(z-w)^2(\ap \up_0^+)^2}\right] \nonumber \\
&\times e^{i\frac{(m \uL(z) + n \uL(w))}{\ap \up_0^+}} \nonumber \\
=&~i\frac{(m-n)}{\ap \up_0^+} \oint_z :\del_z\uL^- e^{i \frac{(m+n)\uL^+(z)}{\ap \up_0^+}}: - \hap \frac{inm^2}{(\ap \up_0^+)^3}:\del_z \uL^+ e^{i \frac{(m+n)\uL^+(z)}{\ap \up_0^+}}:, \nonumber \\
\label{eq:Ahat_Ahat}
}
and,
\eq{
[\widehat {\cal{\uA}}^-_m,C_n] = &-2\left(\frac{in}{4\ap \up_0^+} + \frac{\uV^-}{2}\right)\left(\left[\oint_w \del_w \uL^-,\oint_z \frac{\del^2_z \uL^+}{\del_z \uL^+}\right]e^{i\frac{(m\uL^+(w) + n\uL^+(z))}{\ap \up_0^+}} \right.\nonumber \\
&\left.+ \left[\oint_w \del_w \uL^-,\oint_z e^{i\frac{n \uL^+(z)}{\ap \up_0^+}}\right]e^{i\frac{m \uL^+(w)}{\ap \up_0^+}}\frac{\del^2_z \uL^+}{\del_z \uL^+}\right) \nonumber \\
= &\left(\frac{-in}{2\ap\up_0^+} + \uV^-\right)\left(-\frac{m^2}{2\ap (\up_0^+)^2}\right)\oint_z :\del_z \uL^+ e^{i\frac{(m+n)\uL^+(z)}{\ap \up_0^+}}: \nonumber \\
&+ \left(\frac{-in}{2\ap \up_0^+} + \uV^-\right)\left(\frac{in}{2\up_0^+}\right)\oint_z:\frac{\del^2_z \uL^+}{\del_z \uL^+}e^{i\frac{(m+n)\uL^+(z)}{\ap \up_0^+}}:.
\label{eq:Ahat_Cn}
}
Using \eqref{eq:Ahat_Ahat} and \eqref{eq:Ahat_Cn} in \eqref{eq:AmAm_full}, we get,
\eq{
[\azp {\cal{\uA}}^-_m, &\azp {\cal{\uA}}^-_n] = \oint_z \Bigg[i\frac{(m-n)}{\ap \up_0^+} \oint_z :\del_z\uL^- e^{i \frac{(m+n)\uL^+(z)}{\ap \up_0^+}}: - \hap \frac{inm^2}{(\ap \up_0^+)^3}:\del_z \uL^+ e^{i \frac{(m+n)\uL^+(z)}{\ap \up_0^+}}: \nonumber \\
+ \left(-\frac{in}{2\ap\up_0^+} + \uV^-\right)&\left(\frac{m^2}{2\ap (\up_0^+)^2}\right)\oint_z :\del_z \uL^+ e^{i\frac{(m+n)\uL^+(z)}{\ap \up_0^+}}: -  \left(\frac{-in}{2\ap \up_0^+} +\uV^-\right)\left(\frac{in}{2\up_0^+}\right)\oint_z:\frac{\del^2_z \uL^+}{\del_z \uL^+}e^{i\frac{(m+n)\uL^+(z)}{\ap \up_0^+}}: \nonumber \\ 
 -\left(-\frac{im}{2\ap\up_0^+} + \uV^-\right)&\left(\frac{n^2}{2\ap (\up_0^+)^2}\right)\oint_z :\del_z \uL^+ e^{i\frac{(m+n)\uL^+(z)}{\ap \up_0^+}}: +\left(\frac{-im}{2\ap \up_0^+} +\uV^-\right)\left(\frac{im}{2\up_0^+}\right)\oint_z:\frac{\del^2_z \uL^+}{\del_z \uL^+}e^{i\frac{(m+n)\uL^+(z)}{\ap \up_0^+}}:\Bigg] \nonumber \\
&\implies [\azp {\cal{\uA}}^-_m, \azp {\cal{\uA}}^-_n] = (m-n)\azp {\cal{\uA}}^-_{m+n} + 2m^3 \delta_{m+n,0}
\label{eq:AmAm_result}
}
In writing the last line, we have implicitly used
\eq{
\uV^- (m^2 - n^2) \oint_{z=0}\frac{dz}{2\pi i} \del \uL^+ e^{\frac{i(m+n)\uL^+}{\ap \up_0^+}} \propto \oint_{z=0} d \left[ e^{\frac{i(m+n)\uL^+}{\ap \up_0^+}}\right] = 0
}
We now define the \textit{improved} Brower operator in a linear dilaton background,
\begin{equation}
   {\widetilde    {\cal{\uA}}}^-_m(E)
   =
   {{\cal{\uA}}}^-_m(E)
   -
   \frac{1}{\ualpha_0^+} \tilde{\cL}_m(E)
   -
   \frac{d-2}{24}
   \frac{1}{\ualpha_0^+}\,
   \delta_{m,0} 
   ,
\end{equation}
where we have defined the Sugawara operators (replacing $\ualpha \rightarrow {\cal\uA}$) as
\begin{align}
\tilde{\cL}_m(E)
=
&
\frac{1}{2}
\sum_{j=2}^{D-1} \sum_{l\in \Z}: {\cal{\uA}}^j_l(E)\, {\cal{\uA}}^j_{m-l}(E) : - i\sqrt{\frac{\ap}{2}}(m+1)\uV^- \ualpha_0^+\delta_{m,0}
,
\end{align}
where the second term (from ${\cal \uA}^+_m$) arising due to the linear dilaton background leaves the following commutators (algebra) invariant w.r.t. the flat space case.

The $\tilde{\cal{\uL}}_n$ satisfy the standard Virasoro algebra for a theory with $d-2$ bosons, namely,
\begin{equation}
  [\tilde{\cL}_m,\, \tilde{\cL}_n]
    = 
    (m-n) \tilde{\cL}_{m+n}
    +
    (d-2)
    \frac{1}{12} m(m^2-1)
    \delta_{m+n,0}
    .
\end{equation}
This is easy to check since the ${\cal{\uA}}^i_m$ and $\ualpha^i_m$ satisfy the exact same algebra.
We then observe that,
\eq{
[{\cal{\uA}}^i_n,\widetilde {\cal{\uA}}^-_m] &= [{\cal{\uA}}^i_n,{\cal{\uA}}^-_m] - \frac{1}{\azp}[{\cal{\uA}}^i_n, \tilde{\cL}_m] \nonumber \\
= \frac{n}{\azp}&{\cal\uA}^i_{m+n} - \frac{1}{\azp}\frac{1}{2}\sum_{j=2}^{d-1}\sum_{l \in \mathbb{Z}}[{\cal{\uA}}^i_n, :{\cal{\uA}}^j_{l}{\cal{\uA}}^j_{m-l}:] \nonumber \\
= \frac{n}{\azp}{\cal\uA}^i_{m+n}& - \frac{1}{\azp}\frac{1}{2}(n {\cal{\uA}}^i_{m+n} + n{\cal{\uA}}^i_{m+n}) = 0,
\label{eq:AiAtil_comm}
}
where in reaching the last line, we have used the commutator $[{\cal \uA}^i_m, {\cal \uA}^j_n] = m \delta_{m+n,0}\delta^{ij}$.

We also have,
\eq{
[\azp &{\cal{\uA}}^-_m, \tilde{\cL}_n] = \frac{1}{2}\sum_{j=1}^{d-2} \left(\sum_{p=-\infty}^0[\azp {\cal{\uA}}^-_m, {\cal{\uA}}^j_p {\cal{\uA}}^j_{n-p}] + \sum_{p=1}^\infty[\azp {\cal{\uA}}^-_m, {\cal{\uA}}^j_{n-p}{\cal{\uA}}^j_p]\right) \nonumber\\
= -\frac{1}{2}&\sum_j \left[\sum_{p=-\infty}^0\left(p {\cal{\uA}}^j_{m+p}{\cal{\uA}}^j_{n-p} + (n-p){\cal{\uA}}^j_p{\cal{\uA}}^j_{m+n-p}\right) + \sum_{p=1}^\infty\left((n-p){\cal{\uA}}^j_{m+n-p}{\cal{\uA}}^j_p + p {\cal{\uA}}^j_{n-p}{\cal{\uA}}^j_{m+p}\right)\right] \nonumber \\
&= -\frac{1}{2}\sum_j\left[\sum_{p=-\infty}^0(n-p){\cal{\uA}}^j_{p}{\cal{\uA}}^j_{m+n-p} + \sum_{q=-\infty}^m (q-m){\cal{\uA}}^j_{q}{\cal{\uA}}^j_{m+n-q} \right. \nonumber \\
&~~~\left. + \sum_{p=1}^\infty (n-p){\cal{\uA}}^j_{m+n+p}{\cal{\uA}}^j_{p} + \sum_{q=m+1}^\infty(q-m){\cal{\uA}}^j_{m+n-q}{\cal{\uA}}^j_{q}\right] \nonumber \\
&= -\frac{1}{2}\sum_j\left[\sum_{p=-\infty}^0(n-m){\cal{\uA}}^j_{p}{\cal{\uA}}^j_{m+n-p} + \sum_{q=1}^m (q-m){\cal{\uA}}^j_{q}{\cal{\uA}}^j_{m+n-q} \right. \nonumber \\
&~~~\left.+ \sum_{p=m+1}^\infty (n-m){\cal{\uA}}^j_{m+n+p}{\cal{\uA}}^j_{p} + \sum_{q=1}^m(q-m){\cal{\uA}}^j_{m+n-q}{\cal{\uA}}^j_{q}\right] \nonumber \\
=& (m-n)\frac{1}{2}\sum_j\sum_{l\in \mathbb{Z}}:{\cal{\uA}}^j_{p}{\cal{\uA}}^j_{m+n-p}: - \frac{1}{2}\sum_j\left(\sum_{q=1}^m q(q-m) \delta_{m+n,0}\right),
}
where, the last summation of the last line in obtained from normal ordering the second term in the penultimate line (all others are already normal ordered!). Finally, we can evaluate the summation using,
\eq{
\sum_{q=1}^m q^2 = \frac{1}{6}m(m+1)(2m+1), ~~ \sum_{q=1}^m q = \frac{1}{2}m(m+1),
}
to obtain,
\eq{
[\azp {\cal{\uA}}^-_m,\tilde{\cL}_n]=(m-n) \tilde{\cL}_{m+n} + \frac{d-2}{12}m(m^2-1)\delta_{m+n,0}.
\label{eq:Am_cL_comm}
}
Using \eqref{eq:AiAtil_comm} and \eqref{eq:Am_cL_comm}, we can finally calculate
\eq{
[\widetilde {\cal{\uA}}^-_m, \widetilde {\cal{\uA}}^-_n] 
&= 
[{\cal{\uA}}^-_m,{\cal{\uA}}^-_n] 
- \frac{1}{\azp}[\tilde{\cL}_m,{\cal{\uA}}^-_n] 
- \frac{1}{\azp}[{\cal{\uA}}^-_m,\tilde{\cL}_n] 
+ \frac{1}{(\azp)^2}[\tilde{\cL}_m,\tilde{\cL}_n] 
\nonumber \\
=& 
\frac{(m-n)}{\azp}{\cal{\uA}}^-_{m+n} 
+ \frac{2m^3}{(\azp)^2}\delta_{m+n,0} 
\nonumber\\
&+ 
\frac{1}{(\azp)^2}
\left[(n-m)\tilde{\cL}_{m+n} 
+ \frac{d-2}{12}n(n^2-1)\delta_{m+n,0}\right] \nonumber \\
&- 
\frac{1}{(\azp)^2}
\Bigg[(m-n)\tilde{\cL}_{m+n} 
+ \frac{d-2}{12}m(m^2-1)\delta_{m+n,0}
\Bigg] 
\nonumber\\
&+ \frac{1}{(\azp)^2}
\left[
(m-n)\tilde{\cL}_{m+n} 
+ \frac{d-2}{12}m(m^2-1)\delta_{m+n,0}
\right] 
\nonumber \\
=& \frac{(m-n)}{\azp}
\left[
{\cal{\uA}}^-_{m+n} 
- \frac{1}{\azp}\tilde{\cL}_{m+n}
\right] 
- \frac{d-2}{12(\azp)^2}n\delta_{m+n,0} 
+ \frac{d-26}{12}\frac{n^3}{(\azp)^2}\delta_{m+n,0} 
\nonumber \\
=&\frac{(m-n)}{\azp}
\left[{\cal{\uA}}^-_{m+n} 
- \frac{1}{\azp}\tilde{\cL}_{m+n} 
+\frac{d-2}{24(\azp)^2}\delta_{m+n,0}\right] 
+ \frac{26-d}{12}\frac{m^3}{(\azp)^2}\delta_{m+n,0} 
\nonumber \\
\implies
&
\left[
\azp\widetilde {\cal{\uA}}^-_m, 
\azp\widetilde {\cal{\uA}}^-_n\right] 
= 
(m-n)\azp \widetilde {\cal{\uA}}^-_{m+n} 
+  \frac{26-d}{12}m^3\delta_{m+n,0}.
}
\section{Derivation of hermiticity properties}
\label{app:Derivation of hermiticity properties}
To calculate the Hermitian conjugation of the $\widetilde{A}^-_m$ operators we proceed step-by-step as in the flat space case, using the explicit mode expansion of the string solution. Then,
\eq{
\left[e^{i\frac{m \uL^+(z)}{\ap p_0^+}}\right]^\dag &= e^{\left[i\frac{m \uL^+(z)}{\ap p_0^+}\right]^\dag} =  \exp\left[\frac{im}{\ap p_0^+}\left(\oh x^+_0 - i \ap p^+_0 \ln(z) + i \shap \sum_{n\ne 0} \frac{\alpha^+_n}{n} z^{-n}\right)\right]^\dag \nonumber \\
&=\exp\left[-\frac{im}{\ap p_0^+}\left(\oh x^+_0 + i \ap p^+_0 \ln(\zbar) - i \shap \sum_{n\ne 0} \frac{\alpha^+_{-n}}{n} \zbar^{-n}\right)\right] \nonumber \\
&= \exp\left[-\frac{im}{\ap p_0^+}\left(\oh x^+_0 - i \ap p^+_0 \ln(\frac{1}{\zbar}) + i \shap \sum_{m\ne 0} \frac{\alpha^+_{m}}{m} \left(\frac{1}{\zbar}\right)^{-m}\right)\right] \nonumber \\
\implies &\left[e^{i\frac{m \uL^+(z)}{\ap p_0^+}}\right]^\dag = e^{-\frac{im \uL^+\left(\frac{1}{\zbar}\right)}{\ap p_0^+}},
}
where we have used that $x_0^\mu$ and $p_0^\mu$ are Hermitian (for $\mu \neq -$) and $(\alpha^+_n)^\dag =\alpha^+_{-n}$. We also compute
\eq{
[\del \uL^-(z)]^\dag &= \left[-i\shap\sum_{n\in \mathbb{Z}}\alpha^-_n z^{-n-1}\right]^\dag = i\shap \sum_{n\in \mathbb{Z}} \alpha^-_{-n}\zbar^{-n-1} + \ap \frac{\uV^-}{\zbar}\nonumber \\
&= i\shap \frac{1}{\zbar^2}\sum_{m\in \mathbb{Z}}\alpha^-_m \left(\frac{1}{\zbar}\right)^{-m-1} + \ap \frac{\uV^-}{\zbar}= -\frac{1}{\zbar^2}\left[-i\shap\sum_{m\in \mathbb{Z}}\alpha^-_m \left(\frac{1}{\zbar}\right)^{-m-1} - \ap\frac{\uV^-}{(1/\zbar)}\right] \nonumber \\
\implies& [\del \uL^\mu(z)]^\dag = -\frac{1}{\zbar^2} \del \uchi^\mu\left(\frac{1}{\zbar}\right),
}
where $\uchi^\mu(z) = \uL^\mu(z) + \ap \uV^\mu \ln(z)$, i.e. the $\del \uL^-$ component develops a shift in its momentum zero-mode under H.C. In particular, a key difference in the linear dilaton background is,
\eq{
[\ualpha_0^-]^\dag = \ualpha_0^- - i \sdap \uV^-
\label{eq:ualpha-diff}
}

Similarly, the Hermitian conjugate of the second derivative
\eq{
[\del^2 \uL^-(z)]^\dag &= -i\shap \sum_{n}(n+1)\alpha^-_{-n}\zbar^{-n-2} - i\shap \left(-i \sdap \frac{\uV^-}{\zbar^2}\right)\nonumber\\
=& -i\shap \sum_m (1-m)\alpha^-_m \left(\frac{1}{\zbar}\right)^{-m-2} \frac{1}{\zbar^4} - \ap \frac{\uV^-}{\zbar^2} \nonumber \\
= \frac{1}{\zbar^4}i\shap \sum_m (m+1) &\left(\alpha^-_m - i\sdap \uV^-\delta_{m,0}\right) \left(\frac{1}{\zbar}\right)^{-m-2} - 2 i\shap \sum_m \left(\alpha^-_m - i\sdap \uV^-\delta_{m,0}\right) \left(\frac{1}{\zbar}\right)^{-m-1}\frac{1}{\zbar^{3}} \nonumber \\
\implies& [\del^2 \uL^\mu(z)]^\dag  = \frac{1}{\zbar^4}\del^2\uchi^\mu\left(\frac{1}{\zbar}\right) + 2 \frac{1}{\zbar^{3}} \del \uchi^\mu\left(\frac{1}{\zbar}\right).
}
Finally, we define
\eq{
\hat Q_{l; m}(E) = -\oint \frac{dz}{2\pi i}\frac{1}{z^{l+1}}e^{\frac{im\uL^+\left(z\right)}{\ap \up_0^+}},
}
where the $'-'$ outside fixes the direction of the loop to anti-clockwise after taking the complex conjugate. Then,
\eq{
\left[\hat Q_{l; m}(E) \right]^\dagger &= (\zbar)^2 \oint \frac{d\left(\frac{1}{\zbar}\right)}{2\pi i}\frac{1}{\left(\frac{1}{\zbar}\right)^{-l+1} \left(\frac{1}{\zbar}\right)^{-2}}e^{-\frac{im\uL^+\left(\frac{1}{\zbar}\right)}{\ap p_0^+}} \nonumber \\
&= \oint \frac{d\left(\frac{1}{\zbar}\right)}{2\pi i}\frac{1}{\left(\frac{1}{\zbar}\right)^{-l+1}}e^{-\frac{im\uL^+\left(\frac{1}{\zbar}\right)}{\ap p_0^+}} \nonumber \\
\implies & \left[\hat Q_{l; m}(E) \right]^\dag = \hat Q_{-l;-m}(E).
\label{eq:Q_conj}
}
To compute the Hermitian conjugate of $\widehat {\cal\uA}^-_m$, we first note that the action of H.C on a normal ordered product of two operators $A := \del \uL^-$ and $B:= e^{\frac{im \uL^+}{\ap p_0^+}}$ is given by,
\eq{
(:AB:)^\dag = B^\dag A^\dag - \langle{AB}\rangle^\dag \nonumber \\
=~:A^\dag B^\dag: + [B^\dag,A^\dag].
}
Using the above results, we then observe that,
\eq{
:\left(\del \uL^-(z) e^{\frac{im\uL^+(z)}{\ap p_0^+}}\right):~&=~:\left(-\frac{1}{\zbar^2}\right)\del \uchi^-\left(\frac{1}{\zbar}\right)e^{-\frac{im\uL^+\left(\frac{1}{\zbar}\right)}{\ap p_0^+}}: + \frac{im}{\ap p_0^+}\left(-\frac{1}{\zbar^2}\right)(-i\ap)\frac{\zbar}{2}i e^{-\frac{im\uL^+\left(\frac{1}{\zbar}\right)}{\ap p_0^+}} \nonumber \\
\implies& [\widehat {\cal\uA}^-_m]^\dag = \widehat {\cal\uA}^-_{-m} + \frac{m}{\alpha^+_0}\widehat Q_{0;-m} + i\sdap \uV^- \widehat Q_{0;-m}, 
}
where the term proportional to $\uV^-$ arises from the definition of the shifted string coordinate $\uchi^-$.
Combining the results above, we get
\eq{
[{\cal\uA}^-_m(E)]^\dag
=&
[\widehat {\cal\uA}^-_m(E)]^\dag
+ i\stap \oint \frac{d\zbar}{2\pi
  i}:\left[\left(\frac{im}{4p_0^+}-\ap \frac{\uV^-}{2}\right)\left(-\frac{1}{\zbar^2}\frac{\del^2
    \uL^+}{\del \uL^+}
  -
  \frac{2}{\zbar}\right)\right]e^{-\frac{im\uL^+\left(\frac{1}{\zbar}\right)}{\ap
    p_0^+}}:
\nonumber \\
=
\widehat {\cal\uA}^-_{-m}
+ 
\frac{m}{\alpha^+_0}&\widehat Q_{0;-m}+i\sdap \uV^-\widehat Q_{0;-m}
+ i\stap \oint \frac{d\left(\frac{1}{\zbar}\right)}{2\pi i}:\left(-
\frac{i(-m)}{4 p_0^+}-\ap \frac{\uV^-}{2}\right)\frac{\del^2 \uL^+}{\del
  \uL^+} e^{-\frac{im\uL^+\left(\frac{1}{\zbar}\right)}{\ap
    p_0^+}}:
\nonumber \\
+
i\stap &\oint \frac{d\left(\frac{1}{\zbar}\right)}{2\pi
  i}:\frac{2im}{4 p_0^+}\frac{1}{\left(\frac{1}{\zbar}\right)^1}
e^{-\frac{im\uL^+\left(\frac{1}{\zbar}\right)}{\ap p_0^+}}: + i\stap (\ap \uV^-) \oint \frac{d\left(\frac{1}{\zbar}\right)}{2\pi
  i}\frac{1}{\left(\frac{1}{\zbar}\right)^1}
e^{-\frac{im\uL^+\left(\frac{1}{\zbar}\right)}{\ap p_0^+}}
\nonumber \\
=&
{\cal\uA}^-_{-m}(E)
+ \left(\frac{m}{\alpha^+_0}\widehat Q_{0;-m} - \frac{m}{\alpha^+_0}\widehat Q_{0;-m}\right) + \left(i\sdap \uV^-\widehat Q_{0;-m} - i\sdap \uV^-\widehat Q_{0;-m}\right) \\
\implies& [{\cal\uA}^-_m(E)]^\dag = {\cal\uA}^-_{-m}(E) \\
\implies& [{\cal\uA}^-_{m}(E)]^{\dag\dag} = {\cal\uA}^-_m(E).
}

We also note that the $\widetilde {\cal\uA}^-_m$ follows the same algebra
under Hermitian conjugation as ${\cal\uA}^-_m$. This is evident from its
definition and the properties of the involved quantities derived in
the preceding sections of the Appendix.

\printbibliography

\end{document}